\documentclass{osa-article}

\journal{osajournal}

\articletype{Research Article}

\usepackage{siunitx}
\begin{document}

\title{Large few-layer hexagonal boron nitride flakes for nonlinear optics}

\author{Nils Bernhardt,\authormark{1,2} Sejeong Kim,\authormark{1,4} Johannes E. Fröch,\authormark{1} Simon White,\authormark{1} Ngoc My Hanh Duong,\authormark{1} Zhe He,\authormark{3} Bo Chen,\authormark{3} Jin Liu,\authormark{3} Igor Aharonovich,\authormark{1} and Alexander S. Solntsev\authormark{1,5}}

\address{
    \authormark{1}School of Mathematical and Physical Sciences, University of Technology Sydney, Ultimo NSW 2007, Australia\\
    \authormark{2}Institut für Optik und Atomare Physik, Technische Universität Berlin, 10623 Berlin, Germany\\
    \authormark{3}Sun Yat-sen University, Guangzhou, Guangdong 510275, China\\
    \authormark{4}Sejeong.Kim-1@uts.edu.au\\
    \authormark{5}Alexander.Solntsev@uts.edu.au}




\begin{abstract}
Hexagonal boron nitride (hBN) is a layered dielectric material with a wide range of applications in optics and photonics. In this work, we demonstrate a fabrication method for few-layer hBN flakes with areas up to \SI{5000}{\square\micro\meter}. We show that hBN in this form can be integrated with photonic microstructures: as an example, we use a circular Bragg grating (CBG). The layer quality of the exfoliated hBN flake on a CBG is confirmed by second-harmonic generation (SHG) microscopy. We show that the SHG signal is uniform across the hBN sample outside the CBG and is amplified in the center of the CBG.
\end{abstract}

\label{sec:Introduction}
Hexagonal boron nitride (hBN) is a layered material analogous to graphene~\cite{Romagnoli2018} and transition metal dichalcogenides (TMDs)~\cite{Manzeli2017,Bernhardt2020}, which have sparked significant interest in the last few years as potential platforms for future photonic~\cite{doi:10.1063/1.5120030} and optoelectronic~\cite{C8CS00206A} applications. Compared to graphene and TMDs, however, hBN has a much larger bandgap of about \SI{6}{\electronvolt} and can be considered a true dielectric. While its wide bandgap poses a challenge to utilize it in optoelectronics as an active material, it makes it an ideal platform for pure photonics, including a wide range of applications in quantum optics and nonlinear frequency conversion~\cite{Caldwell2019}. For the former, its wide bandgap makes hBN transparent throughout the visible range, enabling the usage of its atomic defects as high-quality single-photon emitters at room temperature~\cite{Aharonovich2016}. At the same time, hBN also has a strong quadratic nonlinear optical susceptibility $\chi^{(2)}$~\cite{Li2013}, which facilitates 2D materials for a range of traditional nonlinear optical applications such as nonlinear microscopy~\cite{doi:10.1002/adma.201705963}, spectroscopy~\cite{Solntsev2018} and signal processing~\cite{Schiek2012}. For nonlinear optics, a wide bandgap is particularly important, as it guarantees negligible optical losses across the visible spectrum. It opens a possibility of integrating hBN with a range of optical microstructures, including photonic chips, resonators, and metasurfaces, adding quadratic nonlinear optical functionality to platforms with low or zero $\chi^{(2)}$~\cite{J.W.2018}. In this context, transparency for all interacting wavelengths is essential, enabling the utilization of hBN with high quality factor resonators~\cite{Taylor2017} and waveguides relying on phase-matching~\cite{Chen2017} -- both concepts were demonstrated with TMDs, but lack in performance due to inherent material losses.

However, the aforementioned applications require a reliable fabrication method to obtain large and thin hBN flakes of high optical quality and unperturbed crystalline structure. Compared to graphene and TMDs~\cite{Huang2015}, separating layers in hBN is much harder due to stronger bonds. Several methods have been proposed~\cite{Zhang2017,C7TC04933A,Wang2019}; however, the crystalline quality and the suitability of the fabricated samples for nonlinear optics remained an open question. 


In this paper, we demonstrate a reliable method to produce hBN flakes with high optical quality, few-nm thickness, and areas that reach \SI{5000}{\square\micro\meter}, based on a method developed previously for TMDs~\cite{RN385}. We performed a statistical analysis of the produced flakes measuring their thickness using AFM and their lateral size using optical microscopy. We perform the integration of hBN flakes with a circular Bragg grating (CBG) to demonstrate the photonic integration functionality. The flake quality is evidenced by a set of SHG microscopy measurements, therefore proving the suitability for nonlinear optics applications.



\begin{figure}
    \center
    \includegraphics[scale = 0.553]{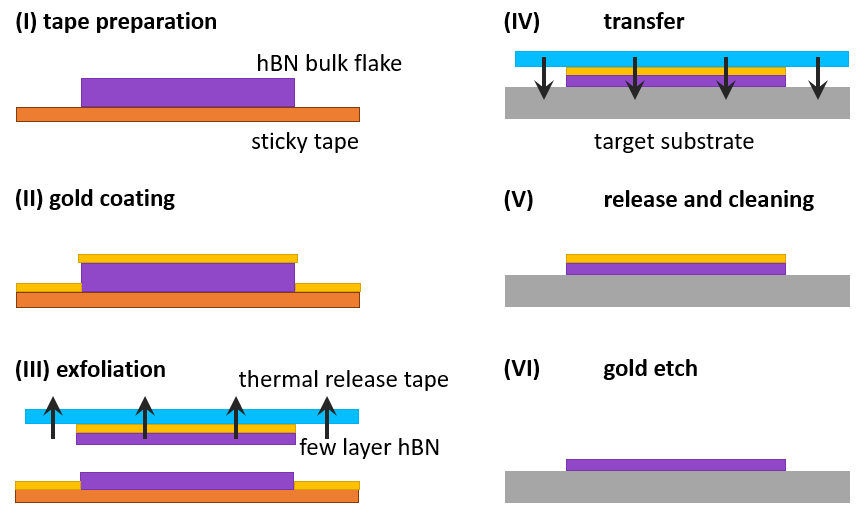} 
    \caption{Schematic of gold-mediated exfoliation of hBN flakes. The process involves six sequential steps and allows fabricating high optical quality hBN samples with a small thickness and large surface area.}
    \label{fig1_exfoliation}
\end{figure}

We obtained a high yield of few-layered hBN flakes by following a previously developed method for TMDs~\cite{RN385}, shown in Fig.~\ref{fig1_exfoliation}. hBN crystals were mechanically exfoliated with sticky tape. Then, the top surface was coated with a \SI{50}{\nano\meter} gold film using sputter deposition. After film deposition, a thermal release tape was used to exfoliate the hBN/gold film. In this case, the sputtered gold film provides a stronger bond to the few topmost layers, thus allowing for a more selective peeling effect using the thermal release tape. Here, the sputter deposition damages the very top surface, causing dangling bonds to better adhere to the sputtered film. The tape was then peeled onto a SiO$_2$ substrate that has undergone additional O$_2$ plasma treatment to increase the surface adhesion. After exfoliation, the tape was released using a hot plate, which left the gold film on top of the hBN intact. After the exfoliation, the substrate was then cleaned using piranha acid, followed by further UV-Ozone treatment. The gold layer was then finally removed using aqua regia.
 
Subsequently, the thin hBN flakes were transferred onto the final target substrate with CBGs using a wet transfer technique. For that, a \SI{500}{\nano\meter} polymethyl methacrylate (PMMA) film was spun onto the SiO$_2$ substrate. Then, the SiO$_2$ substrate was etched in a solution of potassium hydroxide, which would release the PMMA film floating on the surface of the liquid. Another silicon substrate was used to pick up the PMMA film and transfer it to ultra-pure water (repeated three times). Finally, the film was picked up using the target substrate (fused silica) and dried on a hot plate. Subsequently, the PMMA was removed using a warm acetone bath and further cleaned using an ultraviolet ozone treatment.

\begin{figure}
    \center
    \includegraphics[scale = .24]{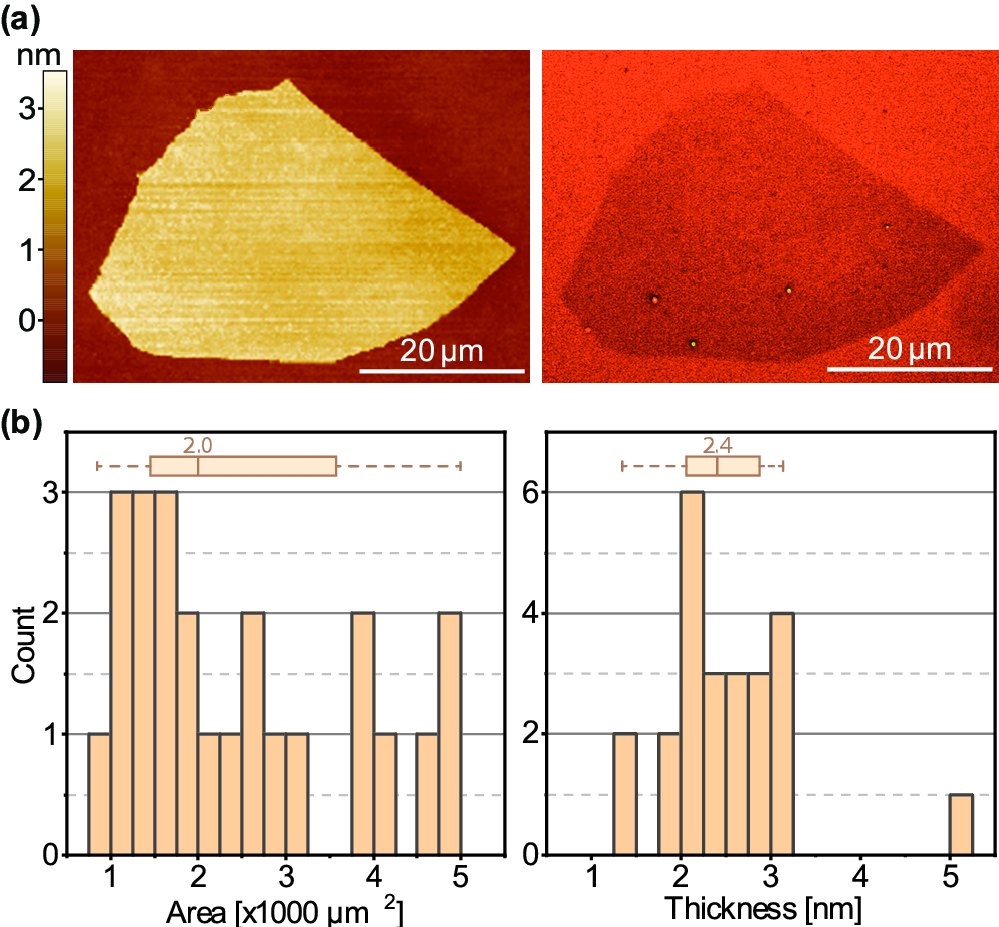}
    \caption{\label{fig:F2_flakes} (a) AFM image (left) and optical image (right) of a  gold-exfoliated hBN flake showing high surface quality and large area. (b) Area and thickness statistics for a set of exfoliated hBN flakes showing the feasibility of this approach for large-area flakes with few-layers thickness. Insets show the respective median values and standard deviations.}
\end{figure}

Microscopy analysis was carried out on a set of hBN flakes to determine the quality of a sample set of hBN flakes following the exfoliation. Fig.~\ref{fig:F2_flakes}a shows side-by-side an atomic force microscope (AFM) image (left) and an optical microscope image (right). From the AFM image, a uniform flat and smooth surface was directly observed. At the same time, the optical image shows a uniform color from the flake surface.

To show the reliability of the fabrication approach, we measured flake size and thickness for a set of 24 flakes, as shown in Fig.~\ref{fig:F2_flakes}b. The thickness of individual flakes was determined from AFM images by averaging the height of an hBN section and relating it to the average surrounding of equivalent surface area size, after leveling the image. The mean flake area is 2,\SI{468}{\square\micro\meter}, for which we assume that the upper limit may be associated with the initial crystalline size due to the flake quality. Hence with larger crystals, larger flakes may be obtainable. The median thickness is \SI{2.405}{\nano\meter}, which corresponds to approximately 7 layers (\SI{3.3}{\angstrom} / layer)~\cite{Kim2012}.


 
After that, we proceeded with photonic integration of hBN flakes on top of a CBG resonator to study nonlinear optical performance. In this work, the use of a CBG is based on both its good emission directionality and intensity enhancement. The CBG structure was fabricated in a \SI{200}{\nano\meter} SiN layer, supported on a \SI{800}{\nano\meter} SiO2 layer on a Si substrate. First, an SiN layer was grown by PECVD (Oxford PlasmaPro System100 ICP180-CVD). After growth, a \SI{400}{\nano\meter} ARP-6200 electron beam resist was spin-cast (4000\,rpm, \SI{60}{\second}) on top of SiN membrane and baked at 150\,$^\circ$C for 3 minutes. The pattern was defined in electron beam resist by electron-beam lithography (Raith Vistec EBPG5000+ \SI{100}{\kilo\volt}) and then transferred into the SiN membrane using a reactive ion etching system (Oxford PlasmaPro System100RIE) with CHF3/O2 gas. Finally, the residual resist was removed by a gentle oxygen plasma RIE process.
The design of the CBG consisted of a wavelength-scale central disk surrounded by an annular periodic grating acting as an antenna optimized for the wavelength of \SI{835}{\nano\meter}.

\begin{figure}
    \center
    \includegraphics[scale = 1]{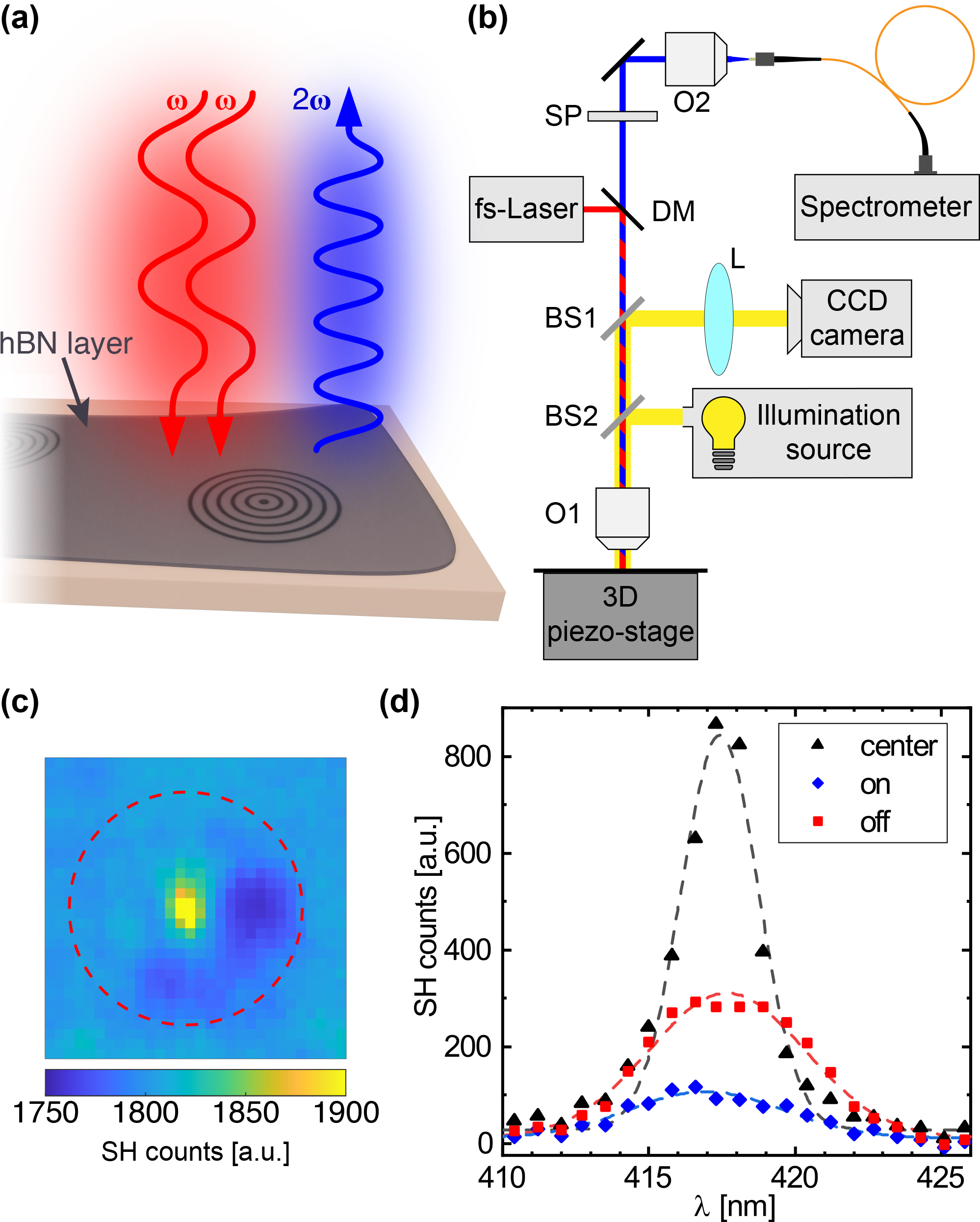}
    \caption{\label{fig:F3_experiment}(a) Schematic representation of the SHG from a few-layered hBN film on a CBG. (b) Simplified schematic of the experimental setup used for the nonlinear measurements of the transferred hBN flake depicted in (a). DM, dichroic mirror; BS1/BS2, beamsplitters; O1/O2, objectives; L, lens; SP, shortpass filter. (c) SHG intensity map of one CBG (marked with a dashed red line) with the hBN flake on top. (d) Measured SHG spectra from hBN in the center (black triangles), on (blue diamonds), and off (red squares) the CBG with respective Gaussian fits (dashed lines).}
\end{figure}

As shown in Fig.~\ref{fig:F3_experiment}a, a 50\,$\times$\,\SI{50}{\micro\meter} hBN flake was transferred onto the CBG structure by use of the stamping method on a lab-built aligned transfer setup consisting of a conventional optical microscope equipped with a long working distance objective (LMPlanFL N 20x), a heating stage, and two micro-manipulators (one to move the sample stage, and the other to move the stamp). The alignment transfer was controlled by observtion through the optical microscope connected to a digital camera. The sample was placed on a heating stage connected to a source meter, and a thermocouple was used to heat up and monitor the temperature of the transfer process. The hBN flake was picked up from the SiO$_2$ substrate at 40\,$^\circ$C, and released onto the structure at 110\,$^\circ$C.

The polydimethylsiloxane (PDMS) stamp was made by mixing 10 parts of Sylgard 184 prepolymer with 1 part of the crosslinking agent. The solution was then spin-coated on a glass slide and cured at 60\,$^\circ$C overnight. Next, the stamp was treated with oxygen plasma for 10 min. Finally, a solution of Polypropylene carbonate (PPC) (15\,$\%$ in anisole, 50\,K, 1500\,rpm) was spin-coated to obtain a \SI{5}{\micro\meter} thick layer on top of the stamp.



To confirm the uniform distribution of layers throughout the whole area of the hBN flake and study the quality of photonic integration with the CBG, we utilized the layer-dependent SHG response~\cite{kim2019second} (see Fig.~\ref{fig:F3_experiment}a). This process requires a non-centrosymmetric crystalline lattice. It only occurs in odd numbers of layers due to destructive interference between the second harmonic generated from even layers and can characterize the non-uniformity of layers in large flakes.

A second harmonic luminescence map was measured using the characterization setup based on a custom scanning confocal microscope, shown in Fig.~\ref{fig:F3_experiment}b. The sample, mounted on a 3D piezo stage, was excited with a pulsed laser at a wavelength of \SI{838}{\nano\meter} (Spectra-Physics MaiTai WB, \SI{80}{\mega\hertz} -- \SI{80}{\femto\second}) using around \SI{30}{\milli\watt} average power. The beam was focused through a 50$\times$ NA\,0.55 objective (O1) onto the hBN flake. The generated second harmonic signal was collected with the same objective; the pump was then filtered using a \SI{750}{\nano\meter} shortpass filter (FES0750) and the signal was directed, via a multi-mode fiber, to an OceanOptics QEPro spectrometer for spectral analysis and 2D SHG mapping.


The sample position was scanned using a piezo stage to measure the 2D map based on the highest SHG intensity for each pixel (Fig.~\ref{fig:F3_experiment}c). The homogeneous second harmonic signal, displayed around the CBG structure (red squares and dashed line in Fig.~\ref{fig:F3_experiment}c), confirms the flake's large size and uniform distribution of layers. The dark sections around the outer edge of the CBG structure are likely due to higher-order diffraction from the grating. At the center of the CBG, second harmonic enhancement can be observed, corresponding to the increased density of states at the resonance combined with the directional out-coupling from the CBG.

Spectrally-resolved second harmonic ($2\omega$) counts are background corrected and fitted with a Gaussian [Fig.~\ref{fig:F3_experiment}(d)]. A \SI{6}{\nano\meter} full width at half maximum (FWHM) signal is observed when the second harmonic is collected at the edge of the grating and off the grating, corresponding to the pump spectrum. At the edge of the grating, the SHG is reduced due to the higher-order diffraction. The grating is designed for the strongest enhancement at 
\SI{836}{\nano\meter} with narrow spectral selectivity. The spectrum shows an enhanced second-harmonic emission around $3\times$ with an FWHM of \SI{3.5}{\nano\meter}, as seen in Fig.~\ref{fig:F3_experiment}d, in agreement with the resonant selectivity of the CBG. These results correspond to what would be expected from a high-quality hBN flake being successfully integrated with the CBG photonic structure.


To conclude, in this work we demonstrate the fabrication of large and thin hBN flakes, perform the study of the reliability of this method, and demonstrate hBN integration with a CBG photonic structure. We perform SHG microscopy that reveals high material and layer quality of the fabricated hBN flakes and successful CBG integration. This work opens hBN for a wide range of applications in nonlinear optics, including microscopy, spectroscopy and signal processing. In this context, hBN can enable strong $\chi^{(2)}$ interactions in well-developed metasurface and photonic chip platforms that would otherwise lack nonlinear quadratic functionality in the visible range. Beyond nonlinear photonics, few-layer hBN flakes are also sought after for quantum tunneling devices~\cite{Parzefall2019} and as thin dielectrics for gating in various configurations~\cite{Gu2019,Binder2019}.

\section*{Disclosures}
The authors declare no conflicts of interest.

\bibliography{article.bib}

\begin{thebibliography}{10}
\newcommand{\enquote}[1]{``#1''}

\bibitem{Romagnoli2018}
M.~Romagnoli, V.~Sorianello, M.~Midrio, F.~H.~L. Koppens, C.~Huyghebaert,
  D.~Neumaier, P.~Galli, W.~Templ, A.~D'Errico, and A.~C. Ferrari,
  \enquote{{Graphene-based integrated photonics for next-generation datacom and
  telecom},} {\protect\JournalTitle{Nature Reviews Materials}} \textbf{3},
  392--414 (2018).

\bibitem{Manzeli2017}
S.~Manzeli, D.~Ovchinnikov, D.~Pasquier, O.~V. Yazyev, and A.~Kis, \enquote{{2D
  transition metal dichalcogenides},} {\protect\JournalTitle{Nature Reviews
  Materials}} \textbf{2}, 17033 (2017).

\bibitem{Bernhardt2020}
N.~Bernhardt, K.~Koshelev, S.~J.~U. White, K.~W.~C. Meng, J.~E. Fröch, S.~Kim,
  T.~T. Tran, D.-Y. Choi, Y.~Kivshar, and A.~S. Solntsev, \enquote{Quasi-bic
  resonant enhancement of second-harmonic generation in ws2 monolayers,}
  {\protect\JournalTitle{Nano Letters}} \textbf{20}, 5309--5314 (2020).

\bibitem{doi:10.1063/1.5120030}
B.~Jia, \enquote{{2D optical materials and the implications for photonics},}
  {\protect\JournalTitle{APL Photonics}} \textbf{4}, 080401 (2019).

\bibitem{C8CS00206A}
V.~W. Brar, M.~C. Sherrott, and D.~Jariwala, \enquote{{Emerging photonic
  architectures in two-dimensional opto-electronics},}
  {\protect\JournalTitle{Chemical Society Reviews}} \textbf{47}, 6824--6844
  (2018).

\bibitem{Caldwell2019}
J.~D. Caldwell, I.~Aharonovich, G.~Cassabois, J.~H. Edgar, B.~Gil, and D.~N.
  Basov, \enquote{{Photonics with hexagonal boron nitride},}
  {\protect\JournalTitle{Nature Reviews Materials}} \textbf{4}, 552--567
  (2019).

\bibitem{Aharonovich2016}
I.~Aharonovich, D.~Englund, and M.~Toth, \enquote{{Solid-state single-photon
  emitters},} {\protect\JournalTitle{Nature Photonics}} \textbf{10}, 631--641
  (2016).

\bibitem{Li2013}
Y.~Li, Y.~Rao, K.~F. Mak, Y.~You, S.~Wang, C.~R. Dean, and T.~F. Heinz,
  \enquote{{Probing Symmetry Properties of Few-Layer MoS 2 and h-BN by Optical
  Second-Harmonic Generation},} {\protect\JournalTitle{Nano Letters}}
  \textbf{13}, 3329--3333 (2013).

\bibitem{doi:10.1002/adma.201705963}
A.~Autere, H.~Jussila, Y.~Dai, Y.~Wang, H.~Lipsanen, and Z.~Sun,
  \enquote{{Nonlinear Optics with 2D Layered Materials},}
  {\protect\JournalTitle{Advanced Materials}} \textbf{30}, 1705963 (2018).

\bibitem{Solntsev2018}
A.~S. Solntsev, P.~Kumar, T.~Pertsch, A.~A. Sukhorukov, and F.~Setzpfandt,
  \enquote{{LiNbO 3 waveguides for integrated SPDC spectroscopy},}
  {\protect\JournalTitle{APL Photonics}} \textbf{3}, 021301 (2018).

\bibitem{Schiek2012}
R.~Schiek, A.~S. Solntsev, and D.~N. Neshev, \enquote{{Temporal dynamics of
  all-optical switching in quadratic nonlinear directional couplers},}
  {\protect\JournalTitle{Applied Physics Letters}} \textbf{100}, 111117 (2012).

\bibitem{J.W.2018}
J.~You, S.~Bongu, Q.~Bao, and N.~Panoiu, \enquote{{Nonlinear optical properties
  and applications of 2D materials: theoretical and experimental aspects},}
  {\protect\JournalTitle{Nanophotonics}} \textbf{8}, 63--97 (2018).

\bibitem{Taylor2017}
T.~Fryett, A.~Zhan, and A.~Majumdar, \enquote{{Cavity nonlinear optics with
  layered materials},} {\protect\JournalTitle{Nanophotonics}} \textbf{7}, 355
  (2017).

\bibitem{Chen2017}
H.~Chen, V.~Corboliou, A.~S. Solntsev, D.-Y. Choi, M.~A. Vincenti,
  D.~de~Ceglia, C.~de~Angelis, Y.~Lu, and D.~N. Neshev, \enquote{{Enhanced
  second-harmonic generation from two-dimensional MoSe2 on a silicon
  waveguide},} {\protect\JournalTitle{Light: Science {\&} Applications}}
  \textbf{6}, e17060--e17060 (2017).

\bibitem{Huang2015}
Y.~Huang, E.~Sutter, N.~N. Shi, J.~Zheng, T.~Yang, D.~Englund, H.-J. Gao, and
  P.~Sutter, \enquote{{Reliable Exfoliation of Large-Area High-Quality Flakes
  of Graphene and Other Two-Dimensional Materials},} {\protect\JournalTitle{ACS
  Nano}} \textbf{9}, 10612--10620 (2015).

\bibitem{Zhang2017}
B.~Zhang, Q.~Wu, H.~Yu, C.~Bulin, H.~Sun, R.~Li, X.~Ge, and R.~Xing,
  \enquote{{High-Efficient Liquid Exfoliation of Boron Nitride Nanosheets Using
  Aqueous Solution of Alkanolamine},} {\protect\JournalTitle{Nanoscale Research
  Letters}} \textbf{12}, 596 (2017).

\bibitem{C7TC04933A}
M.~Sajid, H.~B. Kim, J.~H. Lim, and K.~H. Choi, \enquote{{Liquid-assisted
  exfoliation of 2D hBN flakes and their dispersion in PEO to fabricate highly
  specific and stable linear humidity sensors},} {\protect\JournalTitle{Journal
  of Materials Chemistry C}} \textbf{6}, 1421--1432 (2018).

\bibitem{Wang2019}
N.~Wang, G.~Yang, H.~Wang, C.~Yan, R.~Sun, and C.-P. Wong, \enquote{{A
  universal method for large-yield and high-concentration exfoliation of
  two-dimensional hexagonal boron nitride nanosheets},}
  {\protect\JournalTitle{Materials Today}} \textbf{27}, 33--42 (2019).

\bibitem{RN385}
S.~B. Desai, S.~R. Madhvapathy, M.~Amani, D.~Kiriya, M.~Hettick, M.~Tosun,
  Y.~Zhou, M.~Dubey, J.~W. Ager~III, D.~Chrzan, and A.~Javey,
  \enquote{{Gold-Mediated Exfoliation of Ultralarge Optoelectronically-Perfect
  Monolayers},} {\protect\JournalTitle{Advanced Materials}} \textbf{28},
  4053--4058 (2016).

\bibitem{Kim2012}
K.~K. Kim, A.~Hsu, X.~Jia, S.~M. Kim, Y.~Shi, M.~Hofmann, D.~Nezich, J.~F.
  Rodriguez-Nieva, M.~Dresselhaus, T.~Palacios, and J.~Kong,
  \enquote{{Synthesis of Monolayer Hexagonal Boron Nitride on Cu Foil Using
  Chemical Vapor Deposition},} {\protect\JournalTitle{Nano Letters}}
  \textbf{12}, 161--166 (2012).

\bibitem{kim2019second}
S.~Kim, J.~E. Fr{\"o}ch, A.~Gardner, C.~Li, I.~Aharonovich, and A.~S. Solntsev,
  \enquote{{Second-harmonic generation in multilayer hexagonal boron nitride
  flakes},} {\protect\JournalTitle{Optics letters}} \textbf{44}, 5792--5795
  (2019).

\bibitem{Parzefall2019}
M.~Parzefall, {\'{A}}.~Szab{\'{o}}, T.~Taniguchi, K.~Watanabe, M.~Luisier, and
  L.~Novotny, \enquote{{Light from van der Waals quantum tunneling devices},}
  {\protect\JournalTitle{Nature Communications}} \textbf{10}, 292 (2019).

\bibitem{Gu2019}
J.~Gu, B.~Chakraborty, M.~Khatoniar, and V.~M. Menon, \enquote{{A
  room-temperature polariton light-emitting diode based on monolayer WS2},}
  {\protect\JournalTitle{Nature Nanotechnology}} \textbf{14}, 1024--1028
  (2019).

\bibitem{Binder2019}
J.~Binder, J.~Howarth, F.~Withers, M.~R. Molas, T.~Taniguchi, K.~Watanabe,
  C.~Faugeras, A.~Wysmolek, M.~Danovich, V.~I. Fal'ko, A.~K. Geim, K.~S.
  Novoselov, M.~Potemski, and A.~Kozikov, \enquote{{Upconverted
  electroluminescence via Auger scattering of interlayer excitons in van der
  Waals heterostructures},} {\protect\JournalTitle{Nature Communications}}
  \textbf{10}, 2335 (2019).

\end{thebibliography}

\end{document}